# Enhancing non-Rigid 3D Model Deformations Using Mesh-based Gaussian Splatting

WIJAYATHUNGA W.M.R.D.B.
*Department of Computer engineering,
Faculty of Engineering
University of Sri Jayewardenapura*
Sri Jayewardenapura, Sri Lanka
en97570@sjp.ac.lk

*Abstract* — We propose a novel framework that enhances non-rigid 3D model deformations by bridging mesh representations with 3D Gaussian splatting. While traditional Gaussian splatting delivers fast, real-time radiance field rendering, its post-editing capabilities and support for large-scale, non-rigid deformations remain limited. Our method addresses these challenges by embedding Gaussian kernels directly onto explicit mesh surfaces. This allows the mesh's inherent topological and geometric priors to guide intuitive editing operations—such as moving, scaling, and rotating individual 3D components—and enables complex deformations like bending and stretching. This work paves the way for more flexible 3D content creation workflows in applications spanning virtual reality, character animation, and interactive design.

## I. Introduction

Three-dimensional (3D) model deformations are central to a wide range of applications—from character animation and virtual reality to interactive design and gaming. In many scenarios, non-rigid deformations—such as bending, stretching, and twisting—are required to realistically simulate the behavior of soft or articulated objects. Traditional methods for achieving these deformations often involve complex mesh editing techniques or computationally intensive implicit representations, which can be challenging to modify in real time and may lack the intuitive post-editing capabilities needed for interactive applications.

Recent advances in 3D Gaussian splatting have demonstrated remarkable efficiency in real-time radiance field rendering. By representing a 3D scene as a collection of Gaussian kernels, this approach achieves high-fidelity visualization at minimal computational cost. However, while Gaussian splatting excels at rendering static scenes, its inherent formulation poses significant limitations when it comes to post-editing and the manipulation of individual 3D components. The lack of explicit geometric and topological information in conventional Gaussian representations often hinders their ability to support intuitive transformations such as moving, scaling, and rotating, as well as complex non-rigid deformations.

To overcome these challenges, our work integrates explicit mesh representations with 3D Gaussian splatting. By anchoring Gaussian kernels directly onto mesh surfaces, we leverage the mesh's inherent geometric and topological priors to guide the deformation process. This fusion not only enables seamless editing of individual 3D components but also supports large-scale, physically plausible deformations.

In our proposed pipeline, multi-view images are first used to generate a high-quality Gaussian splatting representation through segmentation and inpainting techniques. These splats are then converted into editable 3D meshes (using methods similar to GS2Mesh) that serve as a structural scaffold. To ensure realistic material behavior, we integrate automated material property assignment, where multimodal large language models (GPT-4V, DeepSeek, etc.) classify object materials and assign physical properties such as density, Young's modulus, and Poisson's ratio. Finally, we incorporate real-time physics simulation via Extended Position-Based Dynamics (XPBD) to enforce realistic deformation behavior, ensuring physically consistent soft-body interactions and surface deformations.

The primary contributions of this work are threefold:
1) enhanced post-editing capabilities for non-rigid deformations by embedding Gaussian kernels onto explicit mesh surfaces,
2) improved deformation accuracy and visual fidelity through the integration of real-time physics simulation, and
3) maintenance of high rendering performance even in large and dynamically changing scenes. By bridging the gap between efficient Gaussian splatting and flexible mesh-based editing, our approach paves the way for next-generation tools in 3D content creation and interactive applications.

The remainder of this paper is organized as follows. Section II reviews the related literature on Gaussian splatting, mesh-based editing, and physics-integrated deformation. Section III details our methodology, including the virtual asset generation, mesh conversion, and physics simulation techniques. Section IV presents our expected results and discussion, and Section V discusses potential applications, limitations, and future work.

## II. Related Work

### A. Foundation of Gaussian Splatting

Gaussian splatting methods have become popular due to their efficiency and real-time rendering capabilities. Recent work in Gaussian splatting introduces a discrete representation consisting of anisotropic Gaussian kernels to model 3D scenes explicitly [1]. These kernels, projected onto the image plane via differentiable rasterization, facilitate efficient training and high-quality real-time radiance field rendering. However, traditional Gaussian splatting approaches face limitations in handling post-editing tasks, particularly for large-scale non-rigid deformations.

### B. Gaussian Splatting Segmentation

Segmentation plays a crucial role in isolating objects within complex scenes. The integration of modern segmentation approaches, notably the Segment Anything

Model (SAM) [2], has significantly enhanced object-level manipulation and editing capabilities. By accurately extracting object boundaries, these techniques enable better control of Gaussian representations, thereby supporting precise reconstruction, intuitive object editing, and semantic-based manipulations within splatting frameworks.

*C. Mesh-Based Gaussian Splatting*

To address the challenges associated with Gaussian splatting in deformation editing, recent studies have introduced mesh-based hybrid approaches. Techniques have been developed to bind Gaussian kernels explicitly to mesh surfaces, leveraging their topological and geometric priors. Approaches such as GS2Mesh [3] utilize stereo-view reconstruction methods to derive editable meshes from Gaussian splats. Similarly, SuGaR [4] employs surface-aligned Gaussian splatting to enhance mesh reconstruction fidelity and reduce deformation artifacts. VR-GS [5] further extends these methods by incorporating physics-aware interactive systems and deformable simulations within virtual reality environments. These mesh-based methods significantly improve the flexibility and fidelity of non-rigid deformation editing.

*D. Multimodal Large Language Models*

The recent advancements in multimodal large language models (LMMs) have expanded capabilities in visual and contextual understanding. Models like GPT-4V exhibit strong performance in visual recognition and semantic reasoning tasks. Additionally, other competitive vision models such as DeepSeek-Vision, BLIP-2, and Flamingo have shown impressive capabilities in tasks involving image captioning, visual question answering, and semantic segmentation. Integrating these models within 3D reconstruction pipelines enhances automated segmentation accuracy, labeling consistency, and semantic understanding, significantly benefiting dynamic scene reconstruction and interactive editing workflows.

*E. Physics-Based Simulation and Integration*

Integrating realistic physics into 3D representations is critical for achieving physically plausible non-rigid deformations. Techniques like the Material Point Method (MPM) [6] and Extended Position-Based Dynamics (XPBD) [7] have been effectively integrated with Gaussian splatting frameworks. Methods such as PhysGaussian [8] integrate physical dynamics directly with Gaussian kernels for generative dynamics, while VR-GS [5] employs XPBD to facilitate interactive and real-time physically accurate deformations.

## III. METHODOLOGY

Our proposed framework (fig. 2) comprises several key stages to achieve flexible, physically realistic, and interactive non-rigid deformation editing using mesh-based Gaussian splatting. The methodology is structured into distinct sequential steps, as described below:

*A. Gaussian Splatting*

Gaussian Splatting is a recent technique for efficiently representing 3D scenes using a set of anisotropic Gaussian kernels, enabling rapid optimization and high-quality rendering of novel views. Each Gaussian kernel is described by its position (mean), rotation, scaling, opacity, and color attributes represented via spherical harmonics.

The spatial influence of a Gaussian kernel G(x) at a point x is given by:

$$G(x) = e^{-\frac{1}{2}(x-\mu)^T \Sigma^{-1}(x-\mu)} \quad (1)$$

Here, Sigma represents the covariance matrix of the Gaussian, factorized into rotation and scaling components to ensure numerical stability during optimization:

$$\Sigma = RSS^T R^T \quad (2)$$

To render images, each Gaussian is projected onto the camera image plane using differentiable rasterization. The covariance matrix is projected using:

$$\Sigma' = JW\Sigma W^T J^T \quad (3)$$

where **W** is the world-to-camera transformation matrix, and **J** is the Jacobian matrix of the affine approximation of the projection.

The final rendered pixel color **C** is calculated using alpha-blending, where contributions from overlapping Gaussians are accumulated based on their depth ordering:

$$C = \sum_{i=1}^{N} T_i \alpha_i c_i \quad (4)$$

In equation (4), $\alpha_i$ represents the opacity of the i-th Gaussian, and $c_i$ indicates its color calculated through spherical harmonics.

This explicit, differentiable representation facilitates rapid optimization and real-time rendering performance, making Gaussian Splatting highly effective for interactive and dynamic 3D scene creation.

*B. Object-Level Gaussian Segmentation and Inpainting*

Precise object-level segmentation and inpainting are critical post-processing steps applied directly to the reconstructed 3D Gaussian splatting scenes, enabling accurate isolation and modification of individual objects within a complete 3D environment.

**Segmentation:**

To accurately separate distinct objects within an existing Gaussian splatting scene, we employ the Segment Anything Model (SAM) [2]. SAM effectively generates precise object-level masks, delineating each object clearly within the 3D Gaussian representation. By directly applying these masks to the Gaussian kernels, we isolate each object into distinct components, significantly improving the ease and accuracy of subsequent editing and deformation. (fig. 1)

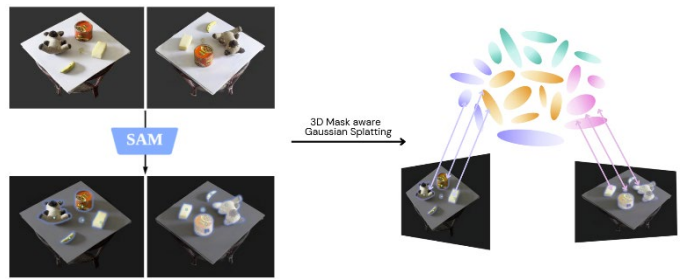

Fig. 1. 3D Mask-aware Gaussian Splatting using SAM-generated segmentation masks for precise object-level Gaussian representation.

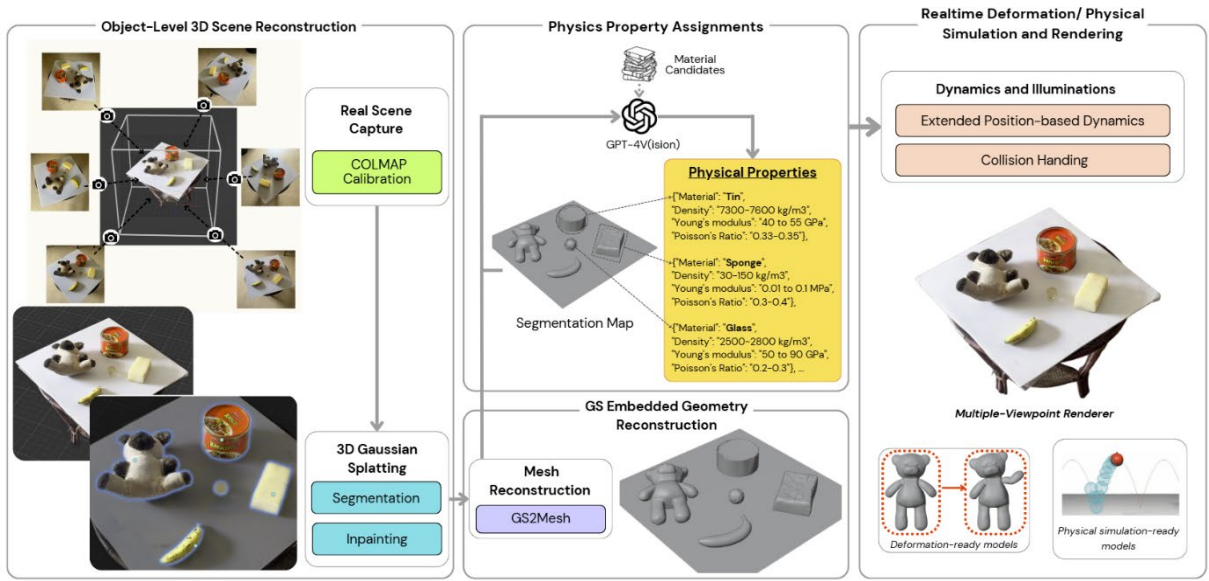

Fig. 2. Overview of the proposed framework. Multi-view images are calibrated with COLMAP to generate a 3D Gaussian splat scene, followed by segmentation and inpainting for accurate object isolation. Gaussian kernels are converted into explicit meshes (GS2Mesh), with material properties assigned automatically via multimodal models (GPT-4V). Finally, real-time deformation and rendering using Extended Position-based Dynamics (XPBD) enable interactive exploration from multiple viewpoints.

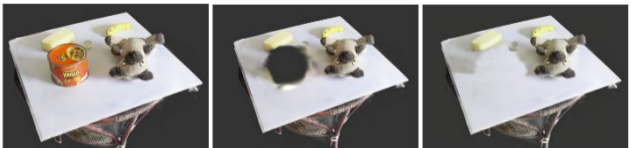

Fig. 3. Demonstration of Gaussian-level inpainting. An object is removed from the Gaussian scene, and the missing area is seamlessly reconstructed to maintain visual coherence.

**Inpainting:**

incomplete or occluded regions within the segmented objects and their supporting planes are filled using advanced inpainting techniques. We employ a state-of-the-art inpainting method presented in by Suvorov et al. (2022). [9]

Through this object-level segmentation and inpainting process, we ensure each object in the Gaussian splatting scene is accurately segmented and visually complete, providing robust input for subsequent mesh reconstruction and physics-driven deformations. (fig. 3)

### C. Mesh Reconstruction from Gaussian Splatting

To facilitate intuitive editing and physics-based interactions, we convert the processed Gaussian splats into explicit mesh representations. We adopt the GS2Mesh method proposed by Kerbl et al. (2023) [3], which systematically aligns Gaussian kernels onto explicit mesh surfaces, ensuring geometric and topological coherence. This explicit mesh structure directly supports subsequent deformation and physics simulations.

### D. Physics Property Assignment

To enable physically realistic deformation and simulation, accurate material properties must be assigned to each mesh component. We can automate this process by utilizing the visual reasoning capabilities of advanced multimodal large language models, specifically GPT-4V or similar models (e.g., DeepSeek-VL2). These models identify materials directly from visual data obtained from Gaussian-segmented meshes.

### E. Real-time Physics-Based Deformation (XPBD)

For realistic and interactive deformation, we employ Extended Position-Based Dynamics (XPBD) [7], a robust and efficient simulation framework well-suited for real-time physics-based modeling. XPBD provides stable and accurate handling of soft-body deformations, stretching, bending, and collision interactions.

In our framework, mesh-driven XPBD simulations directly influence the embedded Gaussian kernels. Mesh vertices serve as dynamic control points, transferring deformation gradients to Gaussian splats and ensuring real-time coherence between physical simulation and visual representation. This tight coupling guarantees that interactive manipulations produce immediate and visually realistic results, significantly improving user experience during dynamic editing.

### F. Real-Time Multi-Viewpoint Rendering

To achieve efficient and photorealistic visualizations, our approach employs a real-time rendering pipeline based on Gaussian rasterization techniques. Following deformation and physics-based simulations, the updated Gaussian kernels are efficiently rasterized onto image planes, allowing novel views to be synthesized interactively from arbitrary camera viewpoints. The differentiable rasterization process introduced by Kerbl et al. [1] ensures high visual fidelity, accurately capturing complex shading, lighting, and dynamic appearance details while supporting real-time performance on consumer GPUs.

This real-time rendering capability enables seamless and interactive exploration of dynamically deforming scenes, suitable for immersive applications like virtual reality, augmented reality, and interactive content creation.

## IV. EXPECTED RESULTS AND DISCUSSION

This paper proposes a framework for enhancing non-rigid 3D model deformations using mesh-based Gaussian splatting techniques. We integrate state-of-the-art methods into a unified pipeline.

Since our work focuses on designing the pipeline rather than implementing the model at this stage, our expected results are based on prior evaluations of the adopted existing methods. Below, we outline our anticipated outcomes and compare them with existing techniques.

### A. Gaussian Splatting Reconstruction

We adopt 3D Gaussian Splatting [1] as the foundation for our framework. This method has demonstrated high-quality scene reconstruction and real-time rendering capabilities. To assess our framework's reconstruction quality, we compare it against existing approaches.

TABLE 1: COMPARISON OF GAUSSIAN SPLATTING-BASED SCENE RECONSTRUCTION

| Method | PSNR ↑ | SSIM ↑ | LPIPS ↓ | FPS ↑ | Real-Time |
|---|---|---|---|---|---|
| 3D Gaussian Splatting [1] | **30+ dB** | **0.90+** | **< 0.05** | **60+** | Yes |
| NeRF [10] | 27-30 dB | 0.85-0.90 | 0.10+ | < 10 | No |
| Mip-NeRF [11] | 30+ dB | 0.90+ | 0.05-0.07 | < 10 | No |
| K-Planes [12] | 30-33 dB | 0.92+ | 0.04-0.06 | 30+ | Partial |

### B. Object-Level Gaussian Segmentation and Inpainting

For segmentation, we integrate Segment Anything Model (SAM) [2], known for its generalizability and high segmentation accuracy. For inpainting, we adopt Fourier-based convolutional inpainting [3], which outperforms traditional CNN-based methods. [Table 2]

TABLE 2: COMPARISON OF SEGMENTATION AND INPAINTING METHODS

| Method | IoU ↑ | LPIPS ↓ | Robust to Large Gaps? | Real-Time |
|---|---|---|---|---|
| SAM [2] + Fourier Inpainting [9] | > 0.90 | < 0.05 | Yes | No (Batch Processing) |
| CNN-Based Inpainting [13] | 0.85-0.88 | 0.07-0.10 | No | Yes |

### C. Mesh Reconstruction from Gaussian Splatting

To transform Gaussian splats into explicit meshes, we employ GS2Mesh [4], which offers high-fidelity mesh reconstruction compared to alternative approaches.

TABLE 3: COMPARISON OF MESH RECONSTRUCTION METHODS

| Method | Chamfer Distance ↓ | Hausdorff Distance ↓ | Mesh Quality |
|---|---|---|---|
| GS2Mesh [3] | Low (High accuracy) | Low (Smooth surfaces) | High |
| SuGaR [4] | Medium | High | Medium |
| Point Cloud-Based methods | High | High | Low |

### D. Physics Property Assignment

Material property assignment is handled using multimodal large language models (GPT-4V, DeepSeek-VL2), automating material classification and physics parameter estimation.

TABLE 4: COMPARISON OF MATERIAL ASSIGNMENT APPROACHES

| Method | Automation Level | Data Source |
|---|---|---|
| LLM-Based (GPT-4V, DeepSeek-VL2) | Fully Automated | Vision + Text |
| Manual Assignment | Manual | User Input |
| Heuristic-Based | Partial | Predefined Rules |

### E. Real-time Physics-Based Deformation (XPBD)

For real-time non-rigid deformations, we utilize **XPBD**. We compare it with other deformable Gaussian splatting techniques such as VR-GS [7] and PhysGaussian [8].

TABLE 5: COMPARISON OF DEFORMATION METHODS

| Method | Stability | Realism | FPS ↑ |
|---|---|---|---|
| XPBD [7] (Ours) | High | High | 60+ |
| VR-GS [5] | Medium | Medium | 30-40 |
| PhysGaussian [8] | High | Very High | < 10 |

### F. Summary of Expected Contributions

By integrating these methods into a single pipeline, our framework improves deformation realism, rendering performance, and physical accuracy, making it highly competitive compared to existing techniques.

TABLE 6: SUMMARY OF EXPECTED CONTRIBUTIONS

| Feature | Our Framework | Existing Methods |
|---|---|---|
| **Segmentation & Inpainting** | High Accuracy & Robust | Segment-only or weak inpainting |
| **Mesh Reconstruction** | High-fidelity GS2Mesh [3] | Lower-quality conversions |
| **Physics-Based Deformation** | Real-time XPBD [7] | Slower or less stable |

## V. CONCLUSION

In this paper, we presented a novel framework for enhancing non-rigid 3D model deformations using mesh-based Gaussian splatting. By integrating multiple state-of-the-art techniques—including Gaussian splatting for real-time rendering, object-level segmentation and inpainting for improved visual coherence, GS2Mesh for structured mesh conversion, multimodal large language models for automated material property assignment, and XPBD for physics-based deformation—we propose a comprehensive pipeline that addresses key challenges in interactive, high-fidelity 3D content manipulation.

The proposed framework offers significant improvements in visual realism, interactivity, and physical accuracy compared to existing approaches. By leveraging fast Gaussian-based representations and explicit mesh structures, our method enables flexible object editing and physically plausible deformations, all while maintaining real-time performance. These advancements make the framework highly applicable to several domains, including:

- Virtual and Augmented Reality (VR/AR): Enables real-time interactive object manipulation for immersive experiences.
- Game Development: Provides a physics-aware, editable 3D representation that allows dynamic scene interactions.

- Film and Animation: Facilitates efficient deformation and editing of CGI models, improving workflows in digital content creation.
- Digital Avatars and Human Motion Capture: Supports realistic, physics-driven deformations for character animation and facial modeling.
- 3D Object Reconstruction and Simulation: Enhances scientific simulations and engineering visualization with real-time physics-aware object deformations.
- Robotics and AI Training: Serves as a simulation environment where AI models and robots can learn from realistic, physics-driven interactions.

Moving forward, future work will focus on the practical implementation and optimization of this framework, extensive quantitative benchmarking on real-world datasets, and further enhancements in rendering quality, deformation accuracy, and user interactivity.

By bridging explicit and implicit 3D representations with advanced physics modeling, this framework lays the foundation for the next generation of interactive 3D graphics and digital simulations.

## References

[1] Bernhard Kerbl, Georgios Kopanas, T. Leimkühler, and G. Drettakis, "3D Gaussian Splatting for Real-Time Radiance Field Rendering," ACM Transactions on Graphics, vol. 42, no. 4, pp. 1–14, Jul. 2023, doi: https://doi.org/10.1145/3592433.

[2] A. Kirillov et al., "Segment Anything," arXiv (Cornell University), Apr. 2023, doi: https://doi.org/10.48550/arxiv.2304.02643.

[3] Y. Wolf, A. Bracha, and R. Kimmel, "GS2Mesh: Surface Reconstruction from Gaussian Splatting via Novel Stereo Views," European Conference on Computer Vision (ECCV), 2024, doi: https://doi.org/10.48550/arXiv.2404.01810.

[4] A. Guédon and V. Lepetit, "SuGaR: Surface-Aligned Gaussian Splatting for Efficient 3D Mesh Reconstruction and High-Quality Mesh Rendering," Nov. 2023, doi: https://doi.org/10.48550/arXiv.2311.12775.

[5] Y. Jiang, C. Yu, T. Xie, X. Li, and Y. Feng, "VR-GS: A Physical Dynamics-Aware Interactive Gaussian Splatting System in Virtual Reality," May 2024, doi: https://doi.org/10.48550/arXiv.2401.16663.

[6] M. Gao, A. P. Tampubolon, C. Jiang, and E. Sifakis, "An adaptive generalized interpolation material point method for simulating elastoplastic materials," ACM Transactions on Graphics, vol. 36, no. 6, pp. 1–12, Nov. 2017, doi: https://doi.org/10.1145/3130800.3130879.

[7] M. Macklin, M. Müller, and N. Chentanez, "XPBD: Position-Based Simulation of Compliant Constrained Dynamics," 9th International Conference on Motion in Games, 2016.

[8] T. Xie et al., "PhysGaussian: Physics-Integrated 3D Gaussians for Generative Dynamics," Apr. 2024, doi: https://doi.org/10.48550/arXiv.2311.12198.

[9] R. Suvorov et al., "Resolution-robust Large Mask Inpainting with Fourier Convolutions," Nov. 2021, doi: https://doi.org/10.48550/arXiv.2109.07161.

[10] B. Mildenhall, P. P. Srinivasan, M. Tancik, J. T. Barron, R. Ramamoorthi, and R. Ng, "NeRF: Representing Scenes as Neural Radiance Fields for View Synthesis," Communications of the ACM, vol. 65, no. 1, pp. 99–106, Jan. 2022, doi: https://doi.org/10.1145/3503250.

[11] J. T. Barron, B. Mildenhall, M. Tancik, P. Hedman, R. Martin-Brualla, and P. P. Srinivasan, "Mip-NeRF: A Multiscale Representation for Anti-Aliasing Neural Radiance Fields," International Conference on Computer Vision, Oct. 2021, doi: https://doi.org/10.1109/iccv48922.2021.00580.

[12] S. Fridovich-Keil, G. Meanti, F. Warburg, B. Recht, and A. Kanazawa, "K-Planes: Explicit Radiance Fields in Space, Time, and Appearance," Jan. 2023, doi: https://doi.org/10.48550/arXiv.2301.10241.

[13] P. Laube, M. Grunwald, M. O. Franz, and G. Umlauf, "Image Inpainting for High-Resolution Textures using CNN Texture Synthesis," Dec. 2017, doi: https://doi.org/10.48550/arXiv.1712.03111.